\documentclass[]{article}

\usepackage{amsmath}
\usepackage{amsmath}
\usepackage{amssymb}
\usepackage{physics}
\usepackage{appendix}
\usepackage{mathtools}

\usepackage{blindtext} 
\usepackage{bookman}
\usepackage[T1]{fontenc} 
\usepackage{microtype} 

\usepackage[english]{babel} 

\usepackage[hmarginratio=1:1,top=32mm,columnsep=20pt]{geometry} 
\usepackage[hang, small,labelfont=bf,up,textfont=rm,up]{caption} 
\usepackage{booktabs} 

\usepackage{lettrine} 

\usepackage{enumitem} 
\setlist[itemize]{noitemsep} 

\usepackage{abstract} 

\usepackage{titlesec} 
\titleformat{\section}[block]{\large\scshape\centering}{\thesection.}{1em}{} 
\titleformat{\subsection}[block]{\large}{\thesubsection.}{1em}{} 

\usepackage{fancyhdr} 
\usepackage{titling} 
\usepackage{cite}
\usepackage{hyperref} 
\hypersetup{
    colorlinks=true,
    linkcolor=blue,
    filecolor=magenta,      
    urlcolor=blue,
    citecolor=blue,
    pdftitle={Overleaf Example},
    pdfpagemode=FullScreen,
    }

\usepackage[dvipsnames]{xcolor}
\usepackage[affil-it]{authblk}
\usepackage{graphicx}

\title{Comments on the gauge dependence of the effective potential and the utility of the  Vilkovisky-DeWitt formalism}
\author{%
\textsc{Daniel W. Collison and Archil Kobakhidze} 
\vspace{0.2cm} \\
\normalsize \itshape
Sydney Consortium for Particle Physics and Cosmology, \\
\normalsize  \itshape
School of Physics, The University of Sydney, NSW 2006, Australia \\ %
}
\date{} 
\renewcommand{\maketitlehookd}{%

\begin{abstract}
 
We provide some additional comments on the long-lived discussions surrounding an effective action and potential plagued by a number of ambiguities. We reinforce the importance of an extra condition on the gauge-fixing function, namely the vanishing of its vacuum expectation value in the absence of external sources, when concluding gauge-independence of the effective action and potential at an extremum. We advocate for the alternative construction of the effective action and potential based on the Vilkovisky-DeWitt approach, and demonstrate its independence from the gauge-fixing parameter. We also exhibit a high-temperature generalisation of this alternative construction in the specific case of the Abelian-Higgs model.
\end{abstract}

}

\begin{document}

\maketitle

\section{Introduction}\label{section-introduction}

The appropriate use of effective potentials in the context of gauge theories has been the subject of ongoing discussions over many decades. Despite the allure of an object as powerful as the effective potential, it is generally neither gauge-invariant, independent of gauge-fixing conditions, nor does it transform as a scalar with respect to field reparametrisations. While certainly not unrelated, these three concepts are nevertheless distinct. 

 From the perspective of the path integral formalism, the effective potential can be found from the more general effective action by evaluating on constant field inputs, and so its lack of gauge-invariance is not particularly surprising in the absence of derivative terms, particularly for gauge fields. Gauge-invariance does not necessarily imply independence of the gauge-fixing conditions used to construct the effective potential however. This distinction was already made explicit, for example, in Kunstatter \cite{Kunstatter:1986qa}, though misuse of terminology may cause confusion. 

 Furthermore, dependence on the gauge-fixing conditions can be subdivided into dependence on the gauge-fixing function and the gauge-fixing parameter. The former selects a particular slice of configuration space in order to avoid counting over gauge-equivalent configurations in the path integral measure. The latter arises from the addition of a BRST-exact term to the classical action. The dependence of the effective action and potential on these quantities is well-documented in a variety of contexts and has been the source of many a headache over the years. The fact that they also do not transform as a scalar with respect to field reparametrisations, though receiving less attention, has also been discussed.  

 The general argument is that because the effective action and potential depend non-trivially on redundancies employed in the construction of the theory from which they arise, such dependencies are considered to relate unphysical information\footnote{However, a quantity which does relate physical information does not automatically mean that it has observable relevance, that is, can be measured in an experiment. This distinction is also referred to by Kummer \cite{Kummer:2001ip}.}, and hence, their interpretation remains problematic. The issue appears to stem from the inclusion of external source terms in the generating functional. These source terms are, in general, neither gauge-invariant nor respect the residual BRST symmetry. In response, various attempts have been made to construct new effective actions and potentials, or argue that certain gauge-fixing conditions or parametrisations for the fields are preferred. The situation is summarised by Andreassen et al. \cite{Andreassen:2014eha}. 

 As also inferred in \cite{Andreassen:2014eha}, there is nothing inherently troubling about objects relating unphysical information depending on the intention behind their usage. One such intention might be to extract quantities that relate only to physical information and move on. However, being familiar with every such quantity appears an exceptionally lofty ambition and most likely unrealistic more generally. Appeals to intuition and to the maintenance of consistency are usually used to pick out quantities which can then be shown to be free of any ambiguities. An important example of this sort includes scattering probabilities derived from S-matrix elements, for which proofs of gauge independence may be found by Kummer \cite{Kummer:2001ip} and an earlier work by Costa and Tonin \cite{Costa:1974rg}. 

 In the context of spontaneous symmetry breaking, the effective potential is of central importance. A promising candidate for a quantity relating to physical information is the value of the effective potential at an extremum. The justification is not without subtleties, however. In this paper, using Nielsen identities, we reinforce the importance of an extra condition that must be imposed on the gauge-fixing function, which was first discussed by de Wit \cite{deWit:1975gzx}. In addition, we also discuss some subtleties surrounding the presence of extremum-producing constant fields, or indeed lack thereof. Not to mention, any practical calculation involving a perturbation expansion must preserve the integrity of any such quantity deemed of physical importance. This is also a delicate and non-trivial matter. 

 Furthermore, we explore an alternative construction of the effective action and potential based on the Vilkovisky-DeWitt formalism. This formalism remoulds the nature of the external source terms, thus attacking the aforementioned issue of dependency on redundancies at its core. It is attractive in that it produces an effective action which is gauge-invariant, independent of gauge-fixing conditions and transformations as a scalar with respect to field reparametrisations without recourse to choosing particular gauge-fixing conditions or parametrisations. We illustrate the independence of such a construction on the gauge-fixing parameter, though there is also an additional condition on the gauge-fixing function that must be satisfied aside from those already discussed in the literature. We also provide an example of the resulting effective potential generalised to high temperature in the context of the Abelian-Higgs model.  

\section{Preliminaries}\label{section-preliminaries}

\subsection{The effective action \& potential}\label{subseceff}

In the interests of providing a brief review of how to obtain the standard effective potential ($V_{eff}$) in the context of the path integral formalism, consider a generic set of fields $\{\varphi^i\}$, with any spacetime or internal indices, as well as spacetime arguments, suppressed for simplicity. The generating functional for this theory is given by 
\begin{align} \label{generatingfunc} 
\mathcal{Z} [J_i, \beta^j;\alpha^k] = \int \mathcal{D} \varphi^i \, \exp( iS[\varphi^i, \beta^j; \alpha^k]+i\int d^4 x \, \varphi^i J_i)
\end{align} 
where $\{\beta^i\}$ and $\{\alpha^i\}$ represent sets of input fields and constant parameters, respectively, and $S$ is the classical action. In addition, the boundary conditions for the fields are chosen carefully to reflect any symmetries imposed onto $S$. The connected generating functional is then given by 
\begin{align}\label{wilfunc}
\mathcal{W}[J_i, \beta^j;\alpha^k] = -i\textrm{ln} \, \mathcal{Z}[J_i, \beta^j;\alpha^k].
\end{align}
Performing a Legendre transform with respect to the set of input source functions  $\{J_i\}$ gives the effective action,
\begin{align}\label{quanteff}
\Gamma[\Phi^i, \beta^j;\alpha^k] = \mathcal{W}[J_i, \beta^j;\alpha^k] - \int d^4x \, \Phi^i J_i
\end{align}
where
\begin{align}\label{meanfield}
\Phi^i =\Phi^i(J_j, \beta^k, \alpha^l)= \frac{\delta\mathcal{W}[J_j, \beta^k; \alpha^l]}{\delta J_i},
\end{align}
are the mean fields (in the presence of input source functions). For practical calculations involving a series expansion, the coefficients of the effective action (or the effective Lagrangian, rather) expressed in powers of the mean field represent the sum of all 1PI diagrams with the corresponding number of external legs for a given power. 

 There is an alternative method for computing the effective action which possesses a number of attractive features and is explicitly presented, for example, by Abbott \cite{Abbott:1981ke}. Consider the modified generating functional
\begin{align} \label{modgenfunc}
\mathcal{Z}' [J_i, \phi^j, \beta^k; \alpha^l]= \int \mathcal{D} \varphi^i \,\, \exp(i S[\varphi^i+\phi^i, \beta^k;\alpha^l]+i\int d^4 x \,\varphi^iJ_i)
\end{align}
where $\{\phi^i\}$ is an additional set of input fields, one for each field in $\{\varphi^i\}$ and with corresponding spacetime and internal indices. Now change functional integration variable by defining $\varphi'^i =\varphi^i + \phi^i$,
\begin{align}\label{shiftvar}
\mathcal{Z}' [J_i, \phi^j, \beta^k; \alpha^l] &= \int \mathcal{D} \varphi'^i \, \exp(iS[\varphi'^i, \beta^k; \alpha^l]+i\int d^4 x \,[\varphi'^iJ_i-\phi^iJ_i]) \nonumber \\ &= \mathcal{Z}[J_i, \beta^k;\alpha^l] \, \exp(-i\int d^4 x \,\phi^iJ_i)
\end{align}
implying
\begin{align}\label{shiftwil} \mathcal{W}'[J_i, \phi^j, \beta^k; \alpha^l]=\mathcal{W}[J_i, \beta^k; \alpha^l] -\int d^4x \,\phi^iJ_i
\end{align}
and thus
\begin{align}\label{shiftmeanfield}
\Phi'^i &=\Phi^i -\phi^i.
\end{align}
Finally, using the previous relations, 
\begin{align}
\Gamma' [\Phi'^i,\phi^j, \beta^k; \alpha^l] &= \mathcal{W}'[J_i,\phi^j, \beta^k; \alpha^l]- \int d^4x \, \Phi'^iJ_i \nonumber \\ &= \mathcal{W}[J_i, \beta^k; \alpha^l]- \int d^4x \, \phi^iJ_i- \int d^4x \, \Phi'^i J_i \nonumber \\  &=\Gamma[\Phi^i, \beta^k; \alpha^l] \nonumber\\ &= \Gamma [\Phi'^i + \phi^i, \beta^k; \alpha^l] \label{shiftgamma}
\end{align}
Setting $\Phi'^i=0$ for all members of the corresponding set, and thus identifying the input field with the mean field, gives $\Gamma'[0,\phi^j, \beta^k; \alpha^l]=\Gamma [\phi^i, \beta^k; \alpha^l]$. That is, in some expansion scheme, the effective action can be obtained by computing all vacuum 1PI graphs in the modified, or shifted, theory and equating the input fields with the mean fields at the end of the calculation. Thus, an advantage of obtaining the effective action in this way is that it appears simpler as only vacuum diagrams are considered.

 $V_{eff}$ can be obtained from the above effective action by simply evaluating on arbitrary spacetime-constant, and hence zero momentum, mean fields. In addition, assume the input fields $\{\beta^i\}$ are set to constant fields for simplicity. Written in momentum space, with  a minus sign and a spacetime volume extracted by convention, 
\begin{align}\label{effpotmomspace}
V_{eff}(\phi^i, \beta^j, \alpha^k)=i \sum_{n=0}^\infty \Gamma^{n}_{a_1\cdots a_n}(p_1=0, \cdots, p_n=0, \beta^k, \alpha^l)\, \phi^{a_1}\cdots\phi^{a_n}
\end{align}
where $\Gamma^{n}_{a_1\cdots a_n}(p_1=0, \cdots, p_n=0, \beta^k, \alpha^l)$ represents the sum of 1PI diagrams with $n$ zero-momentum external legs corresponding to the field species $a_1, \cdots, a_n$. 

 In the case that the expansion scheme of 1PI diagrams involves counting the number of loops, a simple formula for $V_{eff}$ up to and including one-loop contributions is given by  Jackiw \cite{Jackiw:1974cv},
\begin{align}\label{effoneloop}
V_{eff}(\phi^i, \beta^j, \alpha^k)=V_0(\phi^i, \beta^j, \alpha^k)-\frac{i}{2}\int \frac{d^4p}{(2\pi)^4}\ln \det[i\mathcal{D}^{-1}(\phi^i, \beta^j, \alpha^k,p)^l_m]
\end{align}
where $V_0$ is the classical potential and 
\begin{align}\label{effsecdev}
i\mathcal{D}^{-1}(\phi^i, \beta^j, \alpha^k,p)^l_m \coloneqq \int d^4x \,e^{ip\cdot x}\frac{\delta ^2 S[\varphi^n;\beta^j,\alpha^k]}{\delta \varphi_l(x) \delta \varphi^m(0)}
\end{align}
where the RHS is evaluated on the set $\{\phi^i\}$ of spacetime-constant fields.

 To restrict $V_{eff}$ to a subset of constant mean fields, simply set the rest of the fields to zero in $V_{eff}$ where an expression has already been obtained. Alternatively, when working first with the shifted theory, only shift the fields whose associated mean field is desired to feature in $V_{eff}$. Such a restriction must be done with some care, however, as the discussion at the conclusion of section \ref{depgfp} below demonstrates.

\subsection{Gauge-fixing structure \& an example}\label{subsecabel}

Use of the path integral formalism for gauge theories necessitates a gauge-fixing procedure. Such a procedure should eliminate the counting over gauge-equivalent field configurations in the path integral measure. The following provides the general set-up for such a procedure and a specific example in the case of the Abelian-Higgs model.

 As in the previous section, consider a set of fields $\{\varphi^i\}$\footnote{These fields are taken to be non-Grassmann for simplicity and with applications to the Higgs $V_{eff}$ in mind.} with the index running over both generic fields transforming under some representation of the chosen gauge group, and the gauge fields. Again, internal and spacetime indices and spacetime arguments are suppressed. The general structure for the classical gauge-fixed action considered in this paper is 
\begin{align}\label{lagran}
S = \int d^4 x\, \mathcal{L} \coloneqq \int d^4x \, \mathcal{L}_{I} + \mathcal{L}_{GF} + \mathcal{L}_{FPG} 
\end{align}
where $\mathcal{L}_I$ represents the gauge-invariant contribution to the action involving the fields and possible parameters. Furthermore,
\begin{gather}
\mathcal{L}_{GF}  \coloneqq \frac{\xi}{2}B^{2}+B\mathcal{F}, \label{gaugefix}
\\ \mathcal{L}_{FPG}  \coloneqq -\bar{c} \, \mathcal{Q}\mathcal{F} \label{ghost}
\end{gather}
with $\xi$ the gauge-fixing parameter (GFP), $B$ the abelian Nakanishi-Lautrup field, $\mathcal{F}$ the gauge-fixing function (GFF), $\bar{c}$ ($c$) the Fadeev-Popov anti-ghost (ghost) fields and $\mathcal{Q}$ the BRST operator satisfying 
\begin{align}
\mathcal{Q} \varphi^i &= R^i(\varphi^j)\, c & \mathcal{Q} c &= -\frac{1}{2}[c,c] \nonumber
\\ 
\mathcal{Q} B &= 0  & \mathcal{Q} \bar{c} &= B, \label{brsttrans}
\end{align}
where $R^i$ is the gauge generator for the $i$th field, which may depend on the fields, as well as other parameters which are suppressed. 

 It is customary to integrate out the non-dynamical $B$ field in the corresponding path integrals for this theory, resulting in the modification 
\begin{align}\label{modgf}
\mathcal{L}_{GF}=-\frac{\mathcal{F}^2}{2\xi}
\end{align}
together with the addition of an unimportant normalisation factor. Thus, a source function corresponding to the $B$ field, for the purposes of constructing the generating functional for this theory, is not given. In this paper, the $B$ field is retained in order to clearly distinguish between the origins of $\mathcal{F}$ and $\xi$. Attempts to integrate it out are shown explicitly.

 One of the simpler examples, which still retains sufficient complexity for many purposes, is the Abelian-Higgs model. The field content includes a single complex scalar field $\varphi$ coupled to a $U(1)$ gauge field $\mathfrak{A}$. The associated gauge-coupling parameter is denoted $g$. The invariant contribution to the action is given by
\begin{gather}
\mathcal{L}_{I} \coloneqq -\frac{1}{4} F_{\mu \nu} F^{\mu \nu} + (D^{\mu} \varphi )^{\dagger} (D_{\mu} \varphi ) + m^2 \varphi ^{\dagger} \varphi -\frac{\lambda}{4}(\varphi^{\dagger} \varphi)^{2}, \label{gaugeinva}
\end{gather}
with
\begin{gather}
D_{\mu} = \partial_\mu -ig \mathfrak{A}_\mu \label{covder}
\\ F_{\mu \nu}  = \partial_\mu \mathfrak{A}_\nu - \partial_\nu \mathfrak{A}_\mu, \label{abelfields}
\end{gather}
specified by $m^2>0$, where there is an asymmetric vacuum solution already at tree level, and to ensure the presence of stable vacuum solutions, $\lambda>0$ is enforced. In the case for which $m^2=0$ and $\lambda>0$, known as the Coleman-Weinberg model, spontaneous symmetry breaking (SSB) can be induced through the inclusion of higher-order loop corrections \cite{Coleman:1973jx}. 

 A typical choice for the gauge-fixing function in this case is $\mathcal{F}=\partial ^\mu \mathfrak{A}_{\mu}$, for which the subsequent choices $\xi=0,1,3$ are most often used and commonly referred to as the Landau, Feynman and Yennie ``gauges'' respectively with the origin of such designations appearing in \cite{Zumino:1959wt} by Zumino.   In these cases, the ghosts decouple from the rest of the fields and simply contribute to the overall normalisation of the path integral. For other choices of $\mathcal{F}$, this is no longer true.

 The (unrenormalised) $V_{eff}$ for this theory, restricted to the scalar components, up to and including one-loop contributions, is given by 
\begin{gather}
V_{eff }(\phi, \phi^\dagger,m^2, \lambda,g, \xi)=m^2\phi^\dagger\phi -\frac{\lambda}{4}(\phi^\dagger \phi)^2-\frac{i}{2}\int \frac{d^4p}{(2\pi)^4} \Bigg[3\ln(p^2-2 g^2\phi^{\dagger} \phi)\nonumber \\ -\ln (p^2-2g^2\xi\phi^{\dagger} \phi) 
+ \ln\bigg[p^2-\bigg(\frac{\lambda}{2}+2 g^2\xi \bigg)\phi^{\dagger} \phi\bigg] + \ln\bigg(p^2-\frac{3\lambda}{2}\phi^{\dagger} \phi\bigg)\Bigg].
\label{effpotfull}
\end{gather}
The GFF used in the calculation,
\begin{align} \label{gaugefixfunc}
\mathcal{F}= \partial^{\mu} \mathfrak{A}_{\mu} -ig\xi (\phi^{\dagger} \varphi- \varphi^{\dagger}\phi),
\end{align}
is chosen such that the contributions from the scalar-gauge interactions vanish. This is similar to the 't Hooft gauge, except for the identity of the fields $\phi$ and $\phi^\dagger$. As presented in \eqref{gaugefixfunc}, $\{\beta_a\}=\{\phi, \phi^\dagger\}$ are initially input fields of the theory in the language of section \ref{subseceff}. Only in the course of evaluating $V_{eff}$ are they equated with the mean fields. In contrast, in the 't Hooft gauge, these fields are interpreted as the mean fields (though also in the absence of sources) from the beginning. This is another advantage of using the shifted theory approach. Mean fields are initially interpreted as input fields that may be referred to in the classical action, thus avoiding the explicit appearance of a quantity present in the object used to calculate it. 

 The result \eqref{effpotfull} can be explicitly checked using the field-shifting method by computing all the vacuum 1PI diagrams in the shifted theory to one-loop, or by taking advantage of \eqref{effoneloop} directly. The two methods differ only by the addition of field-independent terms. 

\section{Physical nature of SSB}\label{gaugedependence}

\subsection{Dependence on the GFP \& Nielsen identities}\label{depgfp}

As stated in the Introduction, $V_{eff}$ in principle ought to give a powerful insight into the nature of SSB within a theory. However, in gauge theories, $V_{eff}$ is typically also dependent on the choice of gauge-fixing conditions. Indeed, for the Abelian-Higgs model, expression \eqref{effpotfull} reveals the explicit dependence of $V_{eff }$ on the GFP $\xi$. What then are the physical implications of $V_{eff }$ in this theory?

 An answer to such a question potentially arises from considering the relationship between $V_{eff }$ and the GFP in the form of a differential equation, first examined by Nielsen \cite{Nielsen:1975fs} and bearing their name. Such a result follows in general by considering Ward-Takahashi identities arising from BRST invariance. This was followed by various discussions and explicit proofs given certain choices for the form of the GFF, for example, by Aitchison and Fraser \cite{Aitchison:1983ns}, Johnston \cite{Johnston:1984sc}, do Nascimento and Bazeia \cite{DoNascimento:1987mn} and Ramaswamy \cite{Ramaswamy:1995np}, to name a few. These authors mainly focus on the case of scalar electrodynamics and various extensions, including the Abelian-Higgs model and high-temperature scenarios, though there is some discussion of applications to more general theories.   

 Here, we outline a proof of the Nielsen identity in the context of a general theory formulated in the language of Section \ref{section-preliminaries} above. The proof also proceeds from the shifted theory, which may make the calculations of some quantities in the resulting expressions easier. We are not aware of such an approach being taken. 

 First, consider the shifted action \begin{align}
S'[\varphi^i, B, c, \bar{c},\phi^j, \beta^k;\xi, \alpha^l]\coloneqq S[\varphi^i+\phi^i,  B, c, \bar{c}, \beta^k;\xi, \alpha^l] \label{shiftedact}
\end{align}
with the form of the RHS functional given by \eqref{lagran} and where $\{\beta^i\}$ and $\{\alpha^i\}$ are additional sets of input fields and parameters, respectively, and $\phi^i$ may be included in the former set such as seen in \eqref{gaugefixfunc}, prior to any shifting. Typically the goal is to examine the contributions of a subset of the total fields to $V_{eff}$. However, all fields (except the ghost and B fields) are shifted here for reasons that are justified further on in this section. Concerning the gauge-fixed portion of the action, define the shifted GFF\footnote{Also allow for the possibility that the GFF depends explicitly on spacetime points or is otherwise Lorentz non-covariant. In addition, the GFF should remain consistent with any boundary conditions.}
\begin{align}
\mathcal{F}'(\varphi^i, \phi^j, \beta^k,\xi, \alpha^l) \coloneqq \mathcal{F}(\varphi^i+\phi^i,\beta^k, \xi, 
\alpha^l). \label{shiftgff}
\end{align}

 To maintain BRST invariance for the shifted theory, modify the field transformations in \eqref{brsttrans} such that $\mathcal{Q}\varphi^i=R^i(\varphi^j+\phi^j)c$ with $\mathcal{Q}\phi^i=0$. Assume that any additional input fields and parameters transform trivially under BRST transformations. 

 Now promote the GFP to a (scalar) field $\xi=\xi(x)$ and allow $\xi$ to transform non-trivially under a BRST transformation $\mathcal{Q}\xi=\chi$ where $\chi$ is a Grassmann field. In the context of deriving Nielsen identities, such an approach was taken by Johnston \cite{Johnston:1984sc} and Del Cima et al. \cite{DelCima:1999gg} based on earlier work by Piguet and Sibold \cite{Piguet:1984js}.  In this case, to maintain full BRST invariance, the shifted Lagrangian reads
\begin{align}\label{modlaggen}
\mathcal{L}'=\mathcal{L}_I'+\frac{\xi}{2}B^2+\frac{1}{2}\chi\bar{c}B+B\mathcal{F}' -\bar{c}\bigg(\frac{\partial \mathcal{F}'}{\partial 
\varphi^i}\mathcal{Q}\varphi^i  +\frac{\partial \mathcal{F}'}{\partial \xi}\chi \bigg)
\end{align}
Under a BRST transformation of all fields, except $\xi$, the variant portion is 
\begin{align}\label{modlagvar}
\mathcal{Q}\mathcal{L}'&=-\frac{1}{2}\chi B^2 -B\frac{\partial \mathcal{F}'}{\partial \xi}\chi+\bar{c}\frac{\partial^2 \mathcal{F}'}{\partial \varphi^i\partial \xi}\mathcal{Q}\varphi^i\chi  \nonumber\\ &=-\chi\frac{\partial \mathcal{L}'}{\partial \xi}.
\end{align}

 The shifted generating functional, $\mathcal{Z}' [J_i, \bar{G}, G, K_j, \phi^k, \beta^l; \xi, \alpha^m]$, is given by
\begin{gather} \label{modgenfuncniel}
\int \mathcal{D} \varphi^i \mathcal{D}B \mathcal{D} c \mathcal{D} \bar{c} \, \exp[i S' +i\int d^4 x \, (\varphi^i J_i+\bar{G}c +\bar{c}G+\mathcal{Q}\varphi^i_N K_i) ]
\end{gather}
where the new (Grassmann) source fields $K_i$ have been included to account for the non-linear BRST transformations of the non-ghost fields, denoted by the subscript $N$. Thus, the Ward-Takahashi identity corresponding to a BRST transformation of the integrated fields only reads
\begin{gather}
\int d^4 x\mathcal{D} \varphi^i \mathcal{D}B \mathcal{D} c \mathcal{D} \bar{c} \,  \bigg( \mathcal{Q}\varphi^iJ_i +\frac{1}{2}\bar{G}[c,c]+ BG - \chi\frac{\partial \mathcal{L}'}{\partial \xi}\bigg)\nonumber \\ \times \exp[i S' +i\int d^4 x \, (\varphi^i J_i+\bar{G}c +\bar{c}G+\mathcal{Q}\varphi^i_N K_i) ] =0.\label{wardshift}
\end{gather}

 Integrating out the $B$ field at this stage amounts to omitting the functional integration over $B$ and replacing every other instance of it with 
\begin{align}\label{b}
B=-\frac{\mathcal{F}'}{\xi}-\frac{1}{2\xi}\chi\bar{c}.
\end{align}
The corresponding mean fields are given by 
\begin{align} \label{legendre} 
\Phi'^i = \frac{\delta \mathcal{W}'}{\delta J_i},\hspace{0.55cm} \mathcal{C}' = \frac{\delta \mathcal{W}'}{\delta \bar{G}}, \hspace{0.55cm} \bar{\mathcal{C}}' = \frac{\delta \mathcal{W}'}{\delta G}
\end{align}
where $\mathcal{W}'=-i\ln \mathcal{Z}'$ and all arguments have been suppressed. Performing a Legendre transformation on the above variables yields the inverse relations
\begin{align}\label{relleg} 
J_i =- \frac{\delta \Gamma'}{\delta \Phi'^i}, \hspace{0.55cm} \bar{G} = \frac{\delta \Gamma'}{\delta \mathcal{C}'}, \hspace{0.55cm} G =- \frac{\delta \Gamma'}{\delta \bar{\mathcal{C}}'}. \end{align}
Since $K_i$ and $\xi$ are not Legendre transformed,
\begin{align} \label{legendretilde}
\frac{\delta \Gamma'}{\delta K_i} = \frac{\delta \mathcal{W}'}{\delta K_i}, \hspace{0.55cm} \frac{\delta \Gamma'}{\delta \xi} = \frac{\delta \mathcal{W}'}{\delta \xi}. \end{align}

 Putting together all these observations to rewrite \eqref{wardshift} gives
\begin{gather}
\int d^4 x \, \bigg\{- R^i( \mathcal{C}')\frac{\delta \Gamma'}{\delta \Phi_L'^i} +\frac{\delta \Gamma'}{\delta K_i}\frac{\delta \Gamma'}{\delta \Phi_N'^i}  +\frac{1}{2}\frac{\delta \Gamma'}{\delta \mathcal{C}'}\expval{[c,c]}+\frac{1}{\xi}\bigg[\expval{\mathcal{F}'}+\frac{1}{2}\chi \bar{\mathcal{C}}' \bigg]\frac{\delta \Gamma'}{\delta \bar{\mathcal{C}}'} -\chi\frac{\delta \Gamma'}{\delta \xi}  \bigg\}   =0\label{newtakadev}
\end{gather}
where the subscript $L$ labels the linear BRST transformations (which thus are only a function of ghost fields, denoted here by $R$), and $\expval{\cdot}$ represents the expectation value of $\cdot$ in the shifted theory in the presence of the sources given in \eqref{modgenfuncniel}. 

 Functionally differentiating \eqref{newtakadev} with respect to $\chi=\chi(z)$, setting $\chi=0$, and integrating over the $z$ variable gives
\begin{gather}
\int d^4zd^4 x \, \bigg\{ - R^i( \mathcal{C}')\frac{\delta^2 \Gamma'}{\delta \chi \delta \Phi_L'^i}   +\frac{\delta^2 \Gamma'}{\delta \chi \delta K_i}\frac{\delta \Gamma'}{ \delta \Phi_N'^i} -  \frac{\delta \Gamma'}{\delta K_i} \frac{\delta^2 \Gamma'}{\delta \chi \delta \Phi_N'^i}+\frac{1}{2}\frac{\delta^2 \Gamma'}{\delta \chi \delta \mathcal{C}'}\expval{[c,c]} -  \frac{1}{2}\frac{\delta \Gamma'}{\delta \mathcal{C}'} \frac{\delta \expval{[c,c]}}{\delta \chi } \nonumber \\+ \frac{1}{\xi}\frac{\delta \expval{\mathcal{F}'}}{\delta \chi}\frac{\delta \Gamma'}{\delta \bar{\mathcal{C}}'}+ \frac{1}{\xi}\expval{\mathcal{F}'}\frac{\delta^2 \Gamma'}{\delta \chi\delta \bar{\mathcal{C}}'} \bigg\} \bigg|_{\chi=0} +\int d^4 x \bigg\{\frac{1}{2\xi}\bar{\mathcal{C}}'\frac{\delta \Gamma'}{\delta \bar{\mathcal{C}}'}-\frac{\delta \Gamma'}{\delta \xi} \bigg\}\bigg|_{\chi=0}=0. \label{newtakadev2double}
\end{gather}
Setting all fields to zero, including $\chi, K_i$ (but excluding the shifting fields $\phi^i$) and noting the relationship between primed and unprimed quantities derived using the original theory gives 
\begin{gather}
\int d^4zd^4 x \, \bigg\{ \frac{\delta^2 \Gamma'}{\delta \chi \delta K_i}\frac{\delta \Gamma}{ \delta \phi_N^i}  - \frac{\delta \Gamma'}{\delta K_i}\frac{\delta^2 \Gamma'}{\delta \chi \delta \Phi_N'^i} +\frac{1}{2}\frac{\delta^2 \Gamma'}{\delta \chi \delta \mathcal{C}'}\expval{[c,c]} -  \frac{1}{2}\frac{\delta \Gamma}{\delta \mathcal{C}} \frac{\delta \expval{[c,c]}}{\delta \chi }  \nonumber \\+ \frac{1}{\xi}\frac{\delta \expval{\mathcal{F}'}}{\delta \chi}\frac{\delta \Gamma}{\delta \bar{\mathcal{C}}}+ \frac{1}{\xi}\expval{\mathcal{F}'}\frac{\delta^2 \Gamma'}{\delta \chi\delta \bar{\mathcal{C}}'}  \bigg \}\bigg|_{\chi=K=K^\dagger=\text{all fields}=0} -\int d^4 x\,  \frac{\delta \Gamma}{\delta \xi}\bigg|_{\mathcal{C}=\bar{\mathcal{C}}=0} =0. \label{almnieldouble}
\end{gather}
The second and third terms in \eqref{almnieldouble} vanish as they involve either a correlation function involving a product of ghost fields (not including anti-ghosts) or a product of ghost fields and some subset of $\{ \varphi^i\}$ which is zero, or ghost mean fields which are set to zero.

 Finally, evaluating on constant mean, input and $\xi$ fields, and transforming to momentum space gives 
\begin{gather}
\bigg\{- \frac{\delta^2 \Gamma'}{\delta \chi(0) \delta K_i(0)}\frac{\partial V_{eff}}{ \partial \phi^i}+  \frac{1}{2}\frac{\partial V_{eff}}{\partial \mathcal{C}} \frac{\delta \expval{[c,c](0)}}{\delta \chi(0) }   - \frac{1}{\xi}\frac{\delta \expval{\mathcal{F}'(0)}}{\delta \chi
(0)}\frac{\partial V_{eff}}{\partial \bar{\mathcal{C}}} \nonumber \\+ \int \frac{d^4p}{(2\pi)^4} \bigg(\frac{1}{\xi}\expval{\mathcal{F}'(p)}\frac{\delta^2 \Gamma'}{\delta \chi(0)\delta \bar{\mathcal{C}}'(-p)}\bigg)\bigg\}\bigg|_{\chi=K=K^\dagger=\text{all fields}=0}  + \frac{\partial V_{eff}}{\partial \xi}\bigg|_{\mathcal{C}=\bar{\mathcal{C}}=0}  =0. \label{almnielgen} \end{gather}

 Consider \eqref{almnielgen} evaluated at an extremum of $V_{eff}$ with respect to the fields, $\phi^i=\phi^i_0$. An extremum in the ghost fields occurs at $\mathcal{C}=\bar{\mathcal{C}}=0$ automatically given boundary conditions which respect the BRST symmetry. Thus, the first three terms vanish, leaving the bottom line. Note that in this situation $\expval{\mathcal{F}'}=\expval{\mathcal{F}}$ due to the vanishing of the source terms.

 As argued by de Wit in \cite{deWit:1975gzx}, aside from being non-singular, another condition on the GFF necessary to give rise to a consistent gauge theory is that its vacuum expectation value must be zero. This can also be seen straightforwardly by considering a Ward-Takahashi identity applied to the 1-point function for $\bar{c}$ in the absence of sources, given the BRST symmetry of the theory. Given these conditions on the GFF, evaluating \eqref{almnielgen} at an extremum of $V_{eff}$ with respect to the fields gives the result
\begin{gather}
 \frac{\partial V_{eff}}{\partial \xi} =0. \label{xiindepstat}
\end{gather}
That is, the value of $V_{eff}$ at such an extremum is independent of $\xi$ and may represent a possible candidate for a physical quantity associated with SSB in this theory. 

 Notice that if $V_{eff}$ is fully restricted to some subset of $\{\phi^i\}$ with the rest set to zero, it may not be the case that $\expval{\mathcal{F}}$ vanishes at an extremum in the non-zero fields only and hence \eqref{xiindepstat} does not necessarily follow. Thus, in order to maintain \eqref{xiindepstat}, it is necessary to include all field contributions, or at least to set $\phi^i=\phi^i_0$ for the undesired fields, when focusing solely on a subset of contributions to $V_{eff}$. Such a consideration does not extend to the ghosts, as their mean field values at an extremum are automatically zero, as stated previously, regardless of the choice of GFF. In the context of the Abelian-Higgs model, the reason that this is not necessary in the calculation of \eqref{effpotfull}, or indeed when using many of the common gauges, is that $A_0=0$ in these cases. This may not be true in the general case. 

 An important point to note here is that the above discussion presupposes that \emph{constant} mean fields which extremise the effective action, and hence $V_{eff}$, exist. This is always true if the boundary conditions are specified by constant fields\footnote{Or the underlying spacetime manifold has no boundary, but this case is ignored here. It is doubtful whether SSB could even occur in such a theory, as there would be no way of producing multiple extremum-producing mean fields. } and the GFF has no explicit spacetime dependence, as then the effective action is equivalent to $V_{eff}$ up to proportionality. Such a result follows from the translation invariance of the entire generating functional. Specifying constant fields on the boundary can be made compatible with the BRST symmetry, so the above derivation maintains validity. However, the extra restriction on the GFF, satisfying the aforementioned consistency conditions, does not represent the most general choice. 

 Given non-constant boundary conditions for the fields or a more general choice of GFF, it may not be true that such constant extremum-producing mean fields exist or that they exist for all field species at once. For example, in the Abelian-Higgs model case, while such fields may exist corresponding to the scalar and ghost fields, allowing the top line of \eqref{almnielgen} to vanish, it may not be so for the gauge field, preventing the vanishing of $\expval{\mathcal{F}}$. In order to preserve a physically relevant quantity then, it seems necessary to consider the full effective action where an expression analogous to \eqref{xiindepstat} holds, apparently limiting the use of $V_{eff}$ in this sense.

 Furthermore, the conclusions made thus far in this section are non-perturbative. As noted by Andreassen et al. in \cite{Andreassen:2014eha}, there are cases where perturbative approximations to the coefficients in \eqref{almnielgen} yield undefined results, rendering any resulting conclusions meaningless. To capture the non-perturbative conclusions, it is thus necessary to devise a consistent perturbation scheme, an example of which is given in \cite{Andreassen:2014eha}.

\subsection{Dependence on the GFF}\label{depgff}

The last section demonstrated that the value of $V_{eff}$ at a (constant) extremum in the fields is independent of the GFP, provided certain conditions hold with respect to the GFF. However, it did not elucidate the possible dependence on the GFF itself.

 $V_{eff}$ is in fact dependent on the choice of GFF, as seen by using, say, $\mathcal{F}=\partial^\mu \mathfrak{A}_\mu$ instead of \eqref{gaugefixfunc} and re-calculating $V_{eff}$. Most notably, the ghost contributions no longer give rise to field-dependent terms in $V_{eff}$, and the scalar-gauge interactions must be considered. Fortunately, it is straightforward to verify that the value of $V_{eff}$ at an extremum in the fields is also independent of the choice of GFF. For example, at an extremum in the fields, following from \eqref{quanteff},
\begin{align}\label{quanteffgffdep}
\Gamma[ \phi^i_0, 0, 0, \beta^j;\xi,\alpha^k] = \mathcal{W}[0,0,0,\beta^j;\xi,\alpha^k].
\end{align}
By construction, the RHS of \eqref{quanteffgffdep} is independent of the choice of GFF. For constant extremum-producing mean fields, the result carries through to $V_{eff}$.

\section{$V_{eff}$ as a scalar: the Vilkovisky-DeWitt formalism}\label{section-Vilkovisky}

\subsection{Preliminary remarks}\label{prelimvilk}

So far, the value of $V_{eff}$ at a constant extremum in the fields appears to be free from any gauge-fixing ambiguities, given boundary conditions which respect the BRST symmetry. Thus, it is a strong candidate for a physical quantity associated with SSB within the Abelian-Higgs model as suggested previously. Setting aside any discussions involving gravity, the physical quantity is really the difference in values between two extrema. However, it may not be the case that there exist constant field configurations which produce an extrema in the effective action and hence $V_{eff}$. This severely limits the usefulness of such a quantity in characterising SSB in the general case. 

 There is a rather fundamental point concerning the construction of $V_{eff}$, however, which ultimately drives the discussion around its gauge dependence or lack thereof. In defining $V_{eff}$ and the mean fields through \eqref{quanteff} and \eqref{meanfield} respectively, a particular parametrisation for the fields must be chosen. 

 Earlier in this paper, in the context of the Abelian-Higgs model, a complex parametrisation is chosen for the scalar fields $\phi$ and $\phi^\dagger$ together with each component of the gauge field, $\mathfrak{A}_\mu$, being a real number. Another common choice is instead a real parametrisation for the scalar fields $\phi \propto \phi_1+i\phi_2$ and $\phi^\dagger \propto \phi_1-i\phi_2$. Provided ill-defined points are managed carefully, a polar parametrisation $\phi \propto \rho \exp{i\theta}$, $\phi^\dagger \propto \rho \exp{-i\theta}$ and $\mathfrak{A}_\mu=\mathfrak{B}_\mu+\frac{1}{g}\partial_\mu\theta$ is also a viable choice and leads to the so-called unitary Lagrangian. 

 As with any integration variable, it is permissible to use any viable choice of parametrisation to compute quantities of interest. Once the integration is complete, the result should not depend on such a choice. However, in the context of a generating function such as \eqref{generatingfunc}, while a reparametrisation of the fields is always possible, the source terms no longer reflect those of the new fields alone but some possibly non-trivial combination of them. This is not mysterious, but simply reflects the fact that different parametrisations are better suited to computing different observables as discussed by Ellicott and Toms in \cite{Ellicott:1987ir}. What it does mean, though, is that the effective action, and $V_{eff}$, possesses a non-trivial dependence on the choice of field parametrisation in that it does not transform as a scalar with respect to field reparametrisations. In a perturbative expansion, such as in \eqref{effoneloop}, this is clear considering the presence of second derivatives of a scalar function. 

 Indeed, the source terms are often not gauge-invariant, thus breaking the BRST symmetry of the generating functional. It is not surprising that the effective action may have a dependence on the GFP, for example. At an extremum, however, as stated previously, this dependence is not present, nor is any non-trivial parametrisation dependence, as the troublesome source terms are no longer present. This is not a coincidence. The relationship between gauge and parametrisation dependence is inherently a close one as elucidated by the Vilkovisky-DeWitt construction, which results in an effective action that transforms as a scalar with respect to field reparametrisations.

\subsection{Overview and gauge independence}\label{overvilkcon}

 In summary, the details of which can be found in the original paper \cite{Vilkovisky:1984st}, Vilkovisky's proposal is to make the replacement 
\begin{align}\label{replvilk}
 \varphi^iJ_i \rightarrow [\Phi^i-\sigma^i(\Phi^j, \varphi^k)]J_i
\end{align}
for the problematic source term in \eqref{generatingfunc} where $\{\Phi^i\}$ is the set of mean fields and $\sigma^i$ is an object which transforms as a vector with respect to reparametrisations of the mean fields and a scalar with respect to reparametrisations of the original fields $\varphi^i$. Firstly, the geometry of the field space is taken seriously such that the field configurations in some parametrisation form coordinates on the field space manifold. The general idea is to rephrase field-theoretic concepts in parametrisation-covariant language.

 Now consider a vector field $\sigma^i$ residing on the field space manifold satisfying
\begin{align}\label{sigmavilk}
   \sigma^j \,\nabla_j\, \sigma^i=\sigma^i,
   \end{align}
with the boundary conditions $\sigma^i|_{\Phi^j=\varphi^j}= 0$, and $\det(\nabla_j \, \sigma^i)|_{\Phi^k=\varphi^k}\neq 0$ to ensure that $\sigma^i$ is not zero everywhere in any of its components. That is, $\sigma^i$ is tangent to the geodesic connecting $\varphi^i$ and $\Phi^i$ normalised by an affine parameter. Uniqueness of such a geodesic is ensured provided that $\Phi^i$ is in a suitable neighbourhood of $\varphi^i$.  Evaluating $\sigma^i$ at $\Phi^i$, so that it resides in the tangent space corresponding to the point $\Phi^i$, gives the desired transformation properties. For a more detailed explanation of these particulars see \cite{Kunstatter:1986qa}. In addition, if the connection alluded to in \eqref{sigmavilk} obeys a number of properties, then Vilkovisky argues in \cite{Vilkovisky:1984st} that such a connection must coincide with the Christoffel connection constructed from the field space metric given by the kinetic terms in the classical Lagrangian. 

 Implementing the above construction, the resulting effective action transforms as a scalar with respect to field reparametrisations as promised. Moreover, for gauge theories, it can be shown that this effective action is invariant under gauge transformations of the mean fields themselves and is independent of gauge-fixing conditions\footnote{There is a modification of Vilkovisky's original construction due to DeWitt \cite{DeWitt:1985sg} which proves more robust beyond the 1-loop level amongst other attractive properties. We do not suspect that such a modification alters our conclusions here significantly.}. Two crucial results on which these conclusions depend are
\begin{gather}
\mathcal{R}^k(\varphi^j)\frac{\partial}{\partial \varphi^k}\sigma^i(\Phi^l, \varphi^m)=-\mathcal{R}^i(\Phi^n) \nonumber \\
\mathcal{R}^k(\varphi^j)\,\nabla_k \,\sigma^i(\Phi^l, \varphi^m)=\mathcal{R}^i(\Phi^n), \label{cruvilk}
\end{gather}
where $\mathcal{R}^i$ represents the generator of a gauge transformation of the $i$th (non-ghost) field brought about about a change $\mathcal{F}\mapsto \mathcal{F}+\Delta \mathcal{F}$ in the GFF.

 An explicit proof that the new effective action is gauge-invariant and independent of the choice of GFF is given by, for example, Rebhan in \cite{Rebhan:1986wp}, where the only conditions imposed on the GFF as well as its derivative are those of regularity. Rebhan shows that the independence of the choice of GFF follows as a corollary of gauge invariance. What about dependence on the GFP, which is usually related to the preservation of BRST symmetry?

 Here we give an explicit proof of the GFP-independence of the new effective action and its $V_{eff}$ using a similar approach to that of \ref{depgfp} and argue that, not dissimilar to that case, an extra condition on the GFF is necessary to conclude independence.

 Consider the general model from \ref{subsecabel} with the modification to the generating functional given by \eqref{replvilk} for the appropriate field content, and also promote the GFP $\xi$ to a BRST-variant field as done in \ref{depgfp}. Consider the BRST transformation constructed using the gauge generators in $\{\mathcal{R}^i\}$ from \eqref{cruvilk} and derive the resulting Ward-Takahashi identity
\begin{gather}
\int d^4 x \mathcal{D} \varphi^i \mathcal{D}B \mathcal{D} c \mathcal{D} \bar{c} \, \bigg\{ \bigg[  \mathcal{
R}^i(\Phi^j) c \, J_i  +\frac{1}{2}\bar{G}[c,c]  + BG - \chi\frac{\partial \mathcal{L}}{\partial \xi}\bigg]
\nonumber \\ \times \exp[i\bigg( S +\int d^4 x \, (\Phi^i -\sigma^i)J_i+\bar{G}c +\bar{c}G \bigg)] \bigg\} =0.\label{wardshiftvilk}
\end{gather}
Following in a similar vein as in 
\eqref{newtakadev} gives
\begin{gather}
\int d^4 x \, \bigg\{-\mathcal{
R}^i(\Phi^j) \,\mathcal{C}\frac{\delta \Gamma}{\delta \Phi^i}  +\frac{1}{2}\frac{\delta \Gamma}{\delta \mathcal{C}}\expval{[c,c]} + \frac{1}{\xi}\bigg[\expval{\mathcal{F}}+\frac{1}{2}\chi \bar{\mathcal{C}} \bigg]\frac{\delta \Gamma}{\delta \bar{\mathcal{C}}} -\chi\frac{\delta \Gamma}{\delta \xi}  \bigg\}   =0,\label{newtakadevvilk}
\end{gather}
which, after following the same procedure in order to arrive at \eqref{newtakadev2double}, gives,   
\begin{gather}\label{newtakadev2doublevilk}
\int d^4zd^4 x \, \bigg\{ - \mathcal{R}^i(\Phi^j) \,\mathcal{C} \frac{\delta^2 \Gamma}{\delta \chi \delta \Phi^i} +\frac{1}{2}\frac{\delta^2 \Gamma}{\delta \chi \delta \mathcal{C}}\expval{[c,c]} -  \frac{1}{2}\frac{\delta \Gamma}{\delta \mathcal{C}} \frac{\delta \expval{[c,c]}}{\delta \chi }   + \frac{1}{\xi}\frac{\delta \expval{\mathcal{F}}}{\delta \chi}\frac{\delta \Gamma}{\delta \bar{\mathcal{C}}} + \frac{1}{\xi}\expval{\mathcal{F}}\frac{\delta^2 \Gamma}{\delta \chi\delta \bar{\mathcal{C}}} \bigg\} \bigg|_{\chi=0} \nonumber \\+\int d^4 x \bigg(\frac{1}{2\xi}\bar{\mathcal{C}}\frac{\delta \Gamma}{\delta \bar{\mathcal{C}}}-\frac{\delta \Gamma}{\delta \xi} \bigg)=0.
\end{gather}
Setting the mean fields corresponding to the ghosts to zero (which corresponds to extremum-producing values for the effective action as stated previously) and the rest to constants, including $\xi$, gives
\begin{gather}
\int \frac{d^4p}{(2\pi)^4} \bigg(\frac{1}{\xi}\expval{\mathcal{F}(p)}\frac{\delta^2 \Gamma}{\delta \chi(0)\delta \bar{\mathcal{C}}(-p)}\bigg)\bigg|_{\chi=\mathcal{C}=\bar{\mathcal{C}}=0}  + \frac{\partial V_{eff}}{\partial \xi}\bigg|_{\mathcal{C}=\bar{\mathcal{C}}=0}  =0. \label{almnielgenvilk}
\end{gather}

 We argue that in the case of the Vilkovisky construction, in the absence of ghost source functions, $\expval{\mathcal{F}}=0$. The argument is identical that given in \ref{depgfp}, that is, from the Ward-Takahashi identity arising from considering the 1-point function for $\bar{c}$ and the BRST symmetry, now with the modified action. The result then follows that, with ghost contributions set to zero, $V_{eff}$ is independent of the GFP as required.

\subsection{The resulting $V_{eff}$ for the Abelian-Higgs model}\label{section-result}

As an illustration, the resulting $V_{eff}$ of the Abelian-Higgs model as found by Lin and Chyi \cite{Lin:1998up} is given by
\begin{gather}
V_{eff }(\phi, m^2, \lambda, g)=m^2\phi^2 -\frac{\lambda}{4}\phi^4+\frac{1}{2}\int \frac{d^3p}{(2\pi)^3} \Bigg(\omega_A(\overrightarrow{p}) +\omega_\phi(\overrightarrow{p})+\omega_+(\overrightarrow{p})+\omega_-(\overrightarrow{p})\Bigg) \label{effpotfullvilk}
\end{gather}
up to including 1-loop and evaluated on a real ($\phi^\dagger=\phi$), constant field. Here $\omega_A(\overrightarrow{p})\coloneqq \sqrt{(\overrightarrow{p}^2+2g^2\phi^2)}$, $\omega_\phi(\overrightarrow{p}) \coloneqq \sqrt{(\overrightarrow{p}^2-m^2+\frac{3\lambda}{2}\phi^2)}$ and $\omega_\pm(\overrightarrow{p}) \coloneqq \sqrt{(\overrightarrow{p}^2+m_\pm^2)}$, where $m_\pm^2$ are solutions to the equation $(p^2)^2-(4g^2\phi^2+\frac{\lambda}{2}\phi^2-m^2)p^2+4g^4\phi^4=0$. By construction, this result is insensitive to the choice of gauge conditions and the parametrisation used in its derivation, allowing for choices which maximise the simplicity of the calculations. 

 Generalising to the high temperature limit gives
\begin{gather}
V_{eff }(\phi, m^2, \lambda, g,\beta)=m^2\phi^2 -\frac{\lambda}{4}\phi^4+\sum_i\int \frac{d^3p}{(2\pi)^3} \Bigg(\frac{\omega_i(\overrightarrow{p})}{2} +\frac{1}{\beta}\ln\{1-\exp[-\beta \omega_i(\overrightarrow{p})]\}\Bigg) \label{effpotfullvilkhightemp}
\end{gather}
where $i\in \{A,\phi, +,-\}$.

\section{Summary}\label{section-conclusion}

  In this paper, we reinforced the importance an extra condition that must be imposed on the GFF when concluding gauge-independence of the value of $V_{eff}$ at an extremum using Nielsen identities in conjunction with an extended BRST construction. We also discussed some subtleties surrounding the presence of extremum-producing constant fields in general and questioned the use of $V_{eff}$ in the case where no such constant fields exist. 

 In addition, we explored an alternative definition of the effective action and potential given by the Vilkovisky-DeWitt formalism. Such a construction produces an effective action which is gauge-invariant, independent of gauge-fixing conditions and transformations as a scalar with respect to field reparametrisations. We illustrated the independence of such a construction on the GFP in a generalised setting requiring an additional condition on the GFF. We also provided an example of a high-temperature effective potential resulting from this construction in the context of the Abelian-Higgs model. \\

\noindent {\bf Note added} The publication of this paper was postponed for more than two years. During the preparation of our submission, we became aware of an interesting recent work on the subject~\cite{Balui:2025kat}. The authors argue that in $d = 4$ (and, more generally, in any even dimension $d \neq 2$), the functional determinant that arises upon integrating out the quantum fields does not factorize as is commonly assumed when computing the effective potential within the path integral formalism. According to their analysis, this issue appears to stem from regularization ambiguities; once these are properly addressed, the dependence of the effective potential on the gauge parameter $\xi$ is automatically eliminated. Interestingly, no analogous problem arises in odd dimensions or in $d = 2$. Furthermore, the authors claim that no such ambiguity occurs when the effective action is computed using the heat-kernel method~\cite{Balui:2025kat}.

At present, we do not fully understand the physics behind these technical subtleties, nor how these observations relate to the finite-temperature effective potential, particularly when computed in the imaginary-time formalism (i.e., in three dimensions with Euclidean compact time). Moreover, it appears that the effective action remains non-covariant under non-linear reparametrizations, regardless of the chosen regularization scheme. It would be worthwhile to explore the results of~\cite{Balui:2025kat} in greater depth to clarify these points.

\section*{Acknowledgments}
At various stages of this work, we benefited from valuable discussions with Csaba Balázs and Michael Ramsey-Musolf. This research was partially supported by the Australian Research Council under the Discovery Projects grant DP210101636.



\end{document}